\documentclass[runningheads]{llncs}
\usepackage{graphicx}
\usepackage{url}
\pdfoutput=1
\begin{document}
\title{Boosting Connectivity in Retinal Vessel Segmentation via a Recursive Semantics-Guided Network
%\thanks{This work was supported by National Natural Science Foundation of China (NSFC) under Grant 61772106, Grant 61702078 and Grant 61720106005.}
}
\titlerunning{Boosting Vessel Connectivity via a Recursive Semantics-Guided Network}

\author{Rui Xu\inst{1,2,3}\orcidID{0000-0003-0516-3629} \and
Tiantian Liu\inst{2,3,4} \and \\
Xinchen Ye\inst{1,2,3}
\orcidID{0000-0001-5328-3911} 
\and 
Yen-Wei Chen\inst{5}
}

\authorrunning{R. Xu, T. Liu, X. Ye et al.}


\institute{DUT-RU International School of Information Science \& Engineering, \\ Dalian University of Technology, Dalian, China \and
DUT-RU Co-Research Center of Advanced ICT for Active Life, Dalian, China \\
\email{\{xurui,yexch\}@dlut.edu.cn}
\and
Key Laboratory for Ubiquitous Network and Service Software \\ of Liaoning Province, Dalian, China \and
College of Software, Dalian University of Technology, Dalian, China \and
College of Information Science and Engineering, Ritsumeikan University, Kusatsu, Japan
}

\maketitle              % typeset the header of the contribution

\begin{abstract}

Many deep learning based methods have been proposed for retinal vessel segmentation, 
however few of them focus on the connectivity of segmented vessels, 
which is quite important for a practical computer-aided diagnosis system on retinal images. 
In this paper, we propose an efficient network to address this problem. 
A U-shape network is enhanced by introducing a semantics-guided module, 
which integrates the enriched semantics information to shallow layers 
for guiding the network to explore more powerful features. 
Besides, a recursive refinement iteratively applies the same network over the 
previous segmentation results for progressively boosting the performance 
while increasing no extra network parameters. The carefully designed recursive semantics-guided network 
has been extensively evaluated on several public datasets. 
Experimental results have shown the efficiency of the proposed method.

\end{abstract}

\section{Introduction}
\begin{figure*}[!t]
	\centering
	\includegraphics[width=0.9 \linewidth]{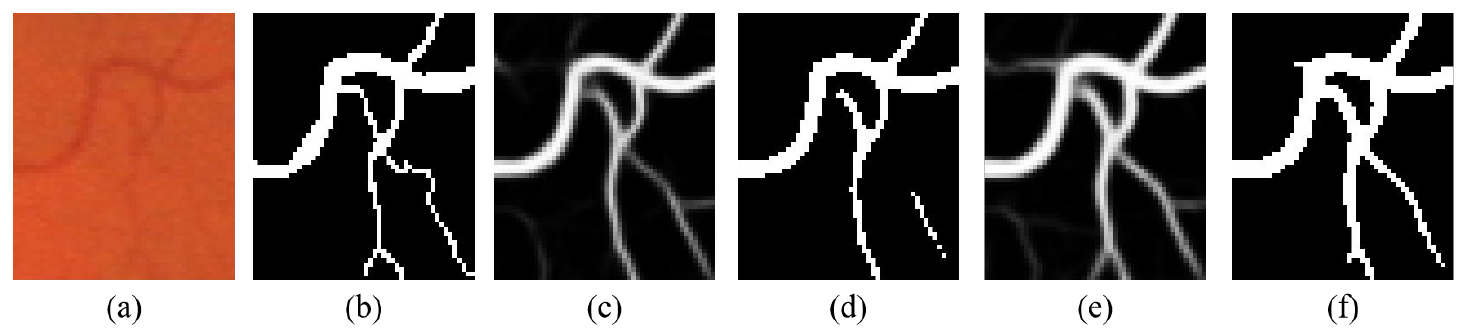}
	\caption{Examples for retinal images and corresponding segmentation results. (a): retinal image. (b): ground truth for vessel segmentation. (c)-(d): predicted probability map and binary segmentation for the work~\cite{oliveira2018retinal}, which achieves relatively high AUC (0.982) but brings breakpoints on segmented vessels. (e)-(f): probability map and binary segmentation for the proposed method, which largely boosts the vessel connectivity. (Otsu's thresholding method~\cite{Otsu2007A} is applied for the probability maps to obtain the binary segmentation results.)}
	\label{fig: example}
	\vspace{-0.5cm}
\end{figure*}

Retinal vessel segmentation is a fundamental and crucial step to develop 
a computer-aided diagnosis (CAD) system on retinal images~\cite{VesReview}.  
Although a lot of deep learning based works have been devoted to precise retinal vessel segmentation, 
few of them have paid attention to the connectivity of segmented vessels. 
In these methods, deep networks are developed to predict dense probability maps indicating how probable 
each pixel belongs to retinal vessels or not, and then retinal vessels are segmented  
by thresholding these maps. Many works have evaluated their methods by calculating some segmentation
metrics, such as the area under the ROC curve (AUC). However, these metrics can not quantify the
topology or connectivity of segmented vessels. Even if high values are achieved for them, 
breakpoints still exist on the binary segmentation results (see Fig.~\ref{fig: example}). 
In this paper, we propose a novel deep network to improve the connectivity on retinal 
vessel segmentation and evaluate it by using some metrics that quantify the topology of segmented vessels.

The U-net architecture is widely adopted in many previous works for retinal vessel segmentation~\cite{fraz2012blood}. 
The original U-net progressively connects high-level layers to shallow layers, 
which makes the semantics information embedded in the high-level layers to be gradually diluted. 
Semantics information is important for retinal vessel segmentation. 
It can not only provide more robust features to boost 
the segmentation of weak vessels while eliminating the effects of abnormal lighting and retinal pathology, 
but also give holistic cues for recognizing whole vessel trees. Thus, semantics information is valuable for  
removing the breakpoints in segmented vessels. 
%There are some works to address this problem of U-net. 
Several previous works try to solve this problem by adding more semantics information into the U-net. 
Wang \textit{et al.} \cite{wang2019dual} design a dual encoding U-Net with a context path 
to capture more semantic information. 
%Ara \textit{et al.} \cite{araujo2019deep} use two 
%cascaded U-Net with a variational auto-encoder to improve the connectivity of blood vessels segmentation. 
Xu \textit{et al.}~\cite{Xu2020} improve U-net by introducing carefully designed semantics and multi-scale 
aggregation blocks. 
%However, these methods make the network more complex and 
%increase a large number of parameters, 
%which heavily raises the difficultly of network training. Different from the previous works, 
%we try a light-weighted manner to improve the U-net that can fully exploit more semantics information 
%to preserve the connectivity on retinal vessel segmentation. 
These works are devoted to designing a complicated network architecture by either 
adding a large number of connections~\cite{Xu2020} or inserting an extra sub-network that is 
relatively large~\cite{wang2019dual}. Different from these efforts, 
we apply several widely-used network block or operation to design a simple but 
efficient module, which can not only fully extract semantics information from high-level layers 
but also guide the network to learn powerful features for better vessel connectivity. 

Besides, refinement is a usual manner to enhance segmentation in literatures and has 
been adopted to improve vessel segmentation in previous works~\cite{araujo2019deep}\cite{wu2018multiscale}. 
Wu \textit{et al.}~\cite{wu2018multiscale} refine the vessel segmentation by using an extra 
multi-scale based network that is cascaded to a preceding network. 
Ara \textit{et al.} \cite{araujo2019deep} have claimed that the connectivity of segmented vessels can be 
improved by stacking another variational auto-encoder based network after a previous one. 
The cascaded manner adopted in these approaches introduces lots of extra network parameters, 
which demands much more labelled data for training. In this paper, 
we adopt a different refinement that uses a single network and recursively refines its output. 
This refinement does not increase extra network parameters while enhancing the connectivity of segmented vessels.
 
Our contribution of this paper can be summarized as follows. 
1) A simple but efficient network is proposed to boost connectivity in retinal vessel segmentation. 
2) A semantics-guided module, which exploits semantics information to guide the network to 
learn more powerful features, is introduced to enhance the capacity of U-net. 
3) A recursive refinement that requires no extra network parameters are exploited to iteratively refine the results.

\begin{figure*}[!t]
	\centering
	\includegraphics[width=0.9\linewidth]{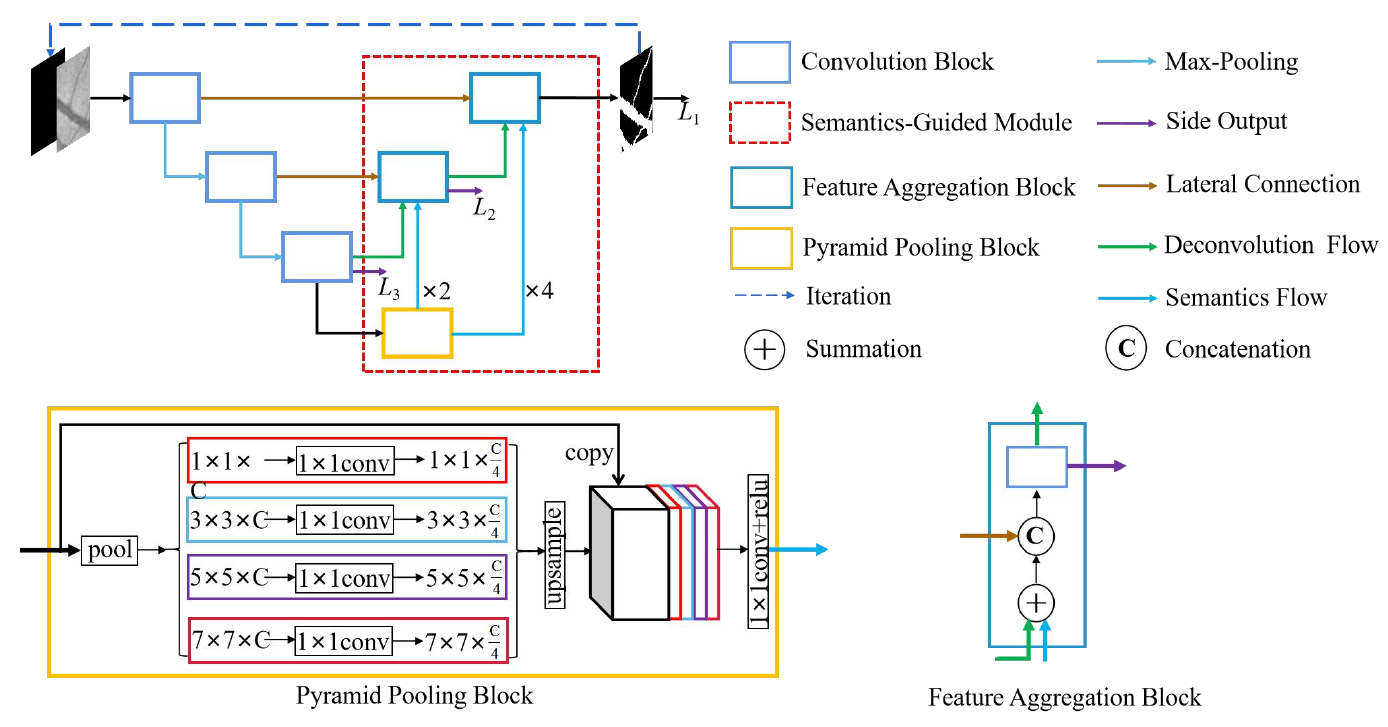}
	\caption{The architecture of the proposed recursive semantics-guided network.}
	\label{fig:network}
	\vspace{-0.5cm}
\end{figure*}

\section{Recursive Semantics-Guided Network}

\subsection{Network Architecture}

We propose a recursive semantics-guided network to enhance the connectivity in retinal vessel segmentation. 
Fig.~\ref{fig:network} illustrates the detailed network architecture, which is designed to address the semantics 
dilution problem of the original U-net. Semantics information is crucial for retinal vessel segmentation. 
It is usually embedded in high-level layers of deep network and less suffered from the effect of some abnormalities, 
such as non-uniform lighting and retinal pathology. It can also provide the holistic cue that is helpful for recognizing 
the whole blood vessel tree. Thus, it is required that semantics information should be fully exploited for 
boosting the vessel segmentation and connectivity. In this paper, we design a semantics-guided module 
that distills the semantics information for guiding the network to produce more powerful and robust features. 
Besides, we adopt a recursive refinement for further boosting the connectivity of segmented vessels. 
Different from the stacked or cascaded refinement, our method re-uses the proposed 
network that iteratively takes the previous results as the input to produce better output. 
We find that this refinement increases no extra network parameters while gradually 
boosting the connectivity of segmented retinal vessels. 

The top part of Fig~\ref{fig:network} shows the whole structure of the proposed network, 
which is based on a 3-layered encoder-decoder architecture. The encoder on each layer 
is a convolutional block comprised by two stacked convolutional layers with $3\times3$ filters 
and the rectified-linear unit (ReLU) activation function. These convolution blocks are serially 
connected by max-pooling that halves feature map sizes to produce hierarchical features in different levels. 
We carefully design the decoder part by introducing a semantics-guided module 
that can fully exploiting semantics information of the deep network. Besides, at the end part on
the 2nd and 3rd layers, there is a side output path that uses upsampling 
and $1\times1$ convolution to obtain a prediction for deep supervision. 
In the following subsections, we describe the semantics-guided module,  
the recursive refinement and network training in details.

%Fig.~\ref{fig:network} gives the proposed method that is a light-weighted but efficient network for blood vessels segmentation. 
%The inputs of the network are $224 \times 224 \times 2$ image patches that are fed into three convolution blocks. Each convolution block is composed by two stacked convolutional layer and rectified-linear unit (ReLU) layer. These convolution blocks are serially connected by max-pooling that halves feature map sizes to produce hierarchical features in different levels. 
%Semantics information can not only provide more robust features to boost the segmentation of weak vessels, but also give holistic cues to recognize whole vessel trees, which are helpful to avoid the breakpoints in segmented vessels. Different from the previous works, we design a semantics-guided module, which exploits semantics information to guide the network to learn more powerful features, to enhance the capacity of U-net. 
%However, some breakpoints still remain in capillary vessels of segmentation results. To improve the connectivity of segmentation results and save parameters, we adopt a different refinement that uses a single network and recursively refines its output.

\subsection{Semantics-Guided Module}

The semantics-guided module is indicated by the red dash block in Fig.~\ref{fig:network}. 
It is mainly comprised by a pyramid pooling block~\cite{PPM} and two feature aggregation blocks. 
The pyramid pooling block is fed with feature maps generated on the 3rd layer. 
As pointed by some previous works~\cite{Pool}\cite{PPM}, the receptive field of a deep network is not wide enough, 
even on the deepest layers. For retinal vessel segmentation, semantics information 
from a wider region can give more holistic cues to identify the global profile of a 
vessel tree for preventing breakpoints. In order to extract more holistic semantics information, 
we connect the pyramid pooling block to the deepest layer. The detailed structure of the 
the pyramid pooling block is given in the left bottom part in Fig.~\ref{fig:network}. 
At first, feature maps are fed into four parallel adaptive pooling layers to produce feature maps with 
the spatial sizes of $1 \times 1$, $3 \times 3$, $5 \times 5$ and $7 \times 7$ respectively. 
Then $1\times1$ convolution followed by upsampling is used to compress channels to be $C/4$ and match their spatial sizes. 
Finally, they are concatenated with the original feature maps and passed to a $1\times 1$ convolution with a ReLU 
activation function to produce the final holistic semantics information, which is provided for the 
feature aggregation blocks. 

The structure of feature aggregation blocks is given in the right-bottom part in Fig.~\ref{fig:network}. 
They exploit semantics information to guide the aggregation of more powerful features. 
Semantic information is inserted via semantics flows that  
utilize upsampling and convolution to match the feature maps coming from a deconvolution flow. 
An element-wise summation is used for fusion and then the results are concatenated with feature maps 
from the lateral connection. Finally, a convolution block consisting of two stacked $3\times3$ convolution is used 
to learn more powerful feature representation. 

\subsection{Recursive Refinement \& Network Training}

The semantics-guided module can improve the connectivity of segmented vessels, however 
the results can be further improved by using a recursive refinement, which gradually 
boosts weak vessels and eliminates break points. In our refinement, predicted vessels 
together with the original image patch are iteratively fed into the same deep network to 
successively obtain a better result. A similar iterative approach is also advocated in previous 
works~\cite{mosinska2018beyond}\cite{Hourglass}, showing that previous results can always be iteratively 
improved when Lipschitz continuity is assumed. 
Besides, this refinement is less demanding of labeled data, since it increases no 
extra network parameters for training. Our recursive refinement can be formulated as 
%$\hat{\mathbf{p}}^{i+1} = f(\mathbf{x}\copyright \hat{\mathbf{p}}^{i}, \mathbf{w})$. 
$\hat{y}^{i+1} = F(x\copyright \hat{y}^{i}) \quad i = 0, 1, \dots, I-1$, 
where $x$ denotes input image patch, $\hat{y}^{i}$ denotes the 
predicted results of a network $F$ in the $i$-th iteration, 
$\copyright$ denotes channel concatenation and $I$ is the total iteration number. 
The refinement is initialized with $\hat{y}^{0} = 0$, which is an empty prediction, and 
the final result is $\hat{y}^{I}$. 

Our network can be trained in an end-to-end manner by minimizing a weighted sum 
of loss functions in all iteration. In each iteration, there are a master $L_{1}$ for predicted results 
and two auxiliary losses ($L_{2}$ and $L_{3}$) for deep supervision. Thus, the total loss in the 
$i$-th iteration can be formulated as $L_{t}^{i}=\sum_{j=1}^{3}{L_{j}^{i}}$. For all of these losses, we adopt 
a weighted binary cross-entropy~\cite{HED}, 
which penalizes more on false negatives than false positives~\cite{araujo2019deep}. 
The final refinement loss can be expressed by $L_{r} = \frac{1}{Z}\sum_{i=1}^{I} i\cdot L_t^i$, 
where $Z$ is the normalization factor $Z = \sum_{i=1}^{I}{i} $. We give more weight for the loss in a later iteration 
to increase numerical stability for network training. In practice, we initially train the network without the refinement, 
which is $I=1$. Then, we increase $I$ to start the training with the refinement. In experiments, 
we set $I$ to be 3.

\begin{figure*}[!t]
	\centering
	\includegraphics[width=0.5 \linewidth]{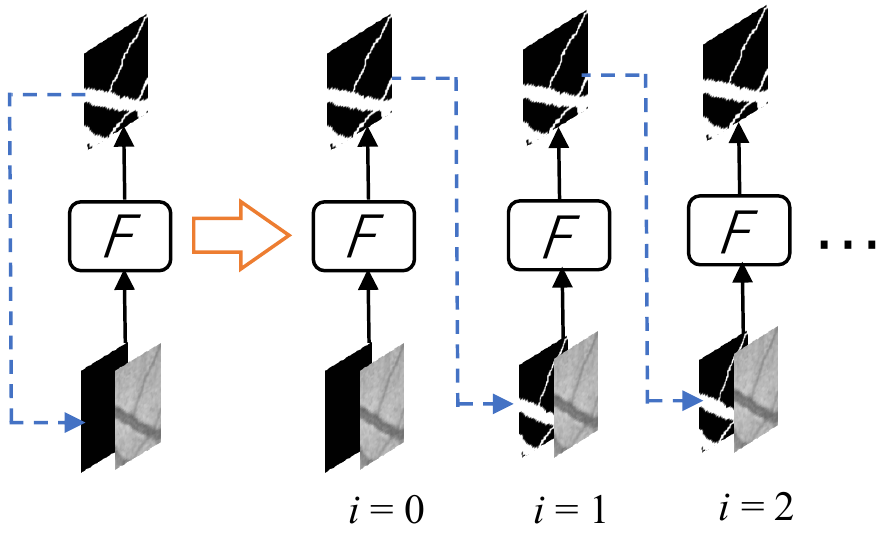}
	\caption{In our recursive refinement, previously predicted results together with the original input are fed into the same network to gradually obtain better results without increasing extra network parameters.}
	\label{fig:refinement}
	\vspace{-0.6cm}
\end{figure*}

\section{Experiments \& Results}

\subsection{Databases \& Evaluation Protocol}
We evaluate the proposed method and other methods by using three publicly
available datasets, which are DRIVE ~\cite{staal2004ridge}, STARE ~\cite{hoover2000locating}, 
and CHASE\_DB1 ~\cite{owen2009measuring}. The DRIVE dataset consists of 40 images, 
7 of which show retinal pathology. The STARE dataset contains 20 images, 
10 of which belong to sick individuals. The CHASE\_DB1 dataset includes 28 images 
that are collected from both eyes of 14 children. The DRIVE dataset is officially divided into 
the training and testing sets that respectively contains 20 images. 
For the STARE and CHASE\_DB1 datasets, we randomly select 10 images for testing, and use the rest for training. 

We evaluate all methods by calculating three widely-used metrics for segmentation and two 
metrics that can quantify the connectivity of segmented vessels. The metrics for segmentation 
include the area under the receiver operating characteristic curve (AUC), sensitivity (SE), and specificity (SP). 
The metrics for quantifying vessel connectivity are previously used for measuring how well the topology of a road 
network is~\cite{TopoMeasures}, and recently adopted to measure the connectivity of 
segmented vessels in~\cite{araujo2019deep}.  
Their calculation requires to randomly select two points which lie both on the ground-truth and binary 
segmentation result. Then, check whether the shortest path between the points has the same length. 
This is repeated many times to record the percentages of correct and infeasible paths. 
Large correct (COR) while low infeasible (INF) percentage indicates good connectivity on segmented vessels. 
In experiments, we find 1000 times repetition per testing image is enough to obtain converged percentages. 
Besides, binary segmentation results are produced apply the Otsu's thresholding method~\cite{Otsu2007A} 
on predicted probability maps for all methods in experiments. 

\subsection{Ablation Study}

\begin{table}[!t]
	\caption{Ablation study of the proposed method with different network configurations for retinal vessel segmentation on the DRIVE dataset. (A smaller value of INF means better performance.)}
	\label{T1}
	\centering
	\includegraphics[width=0.75\linewidth]{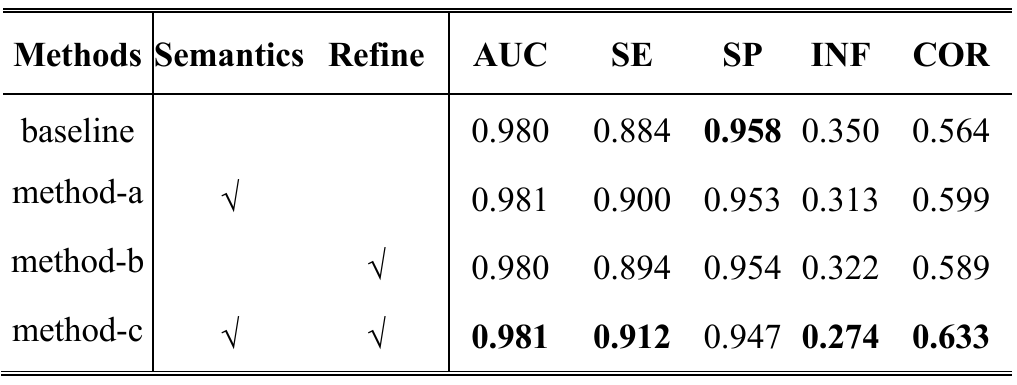}
	\vspace{-0.3cm}
\end{table}

\begin{figure*}[!t]
	\centering
	\includegraphics[width=0.9 \linewidth]{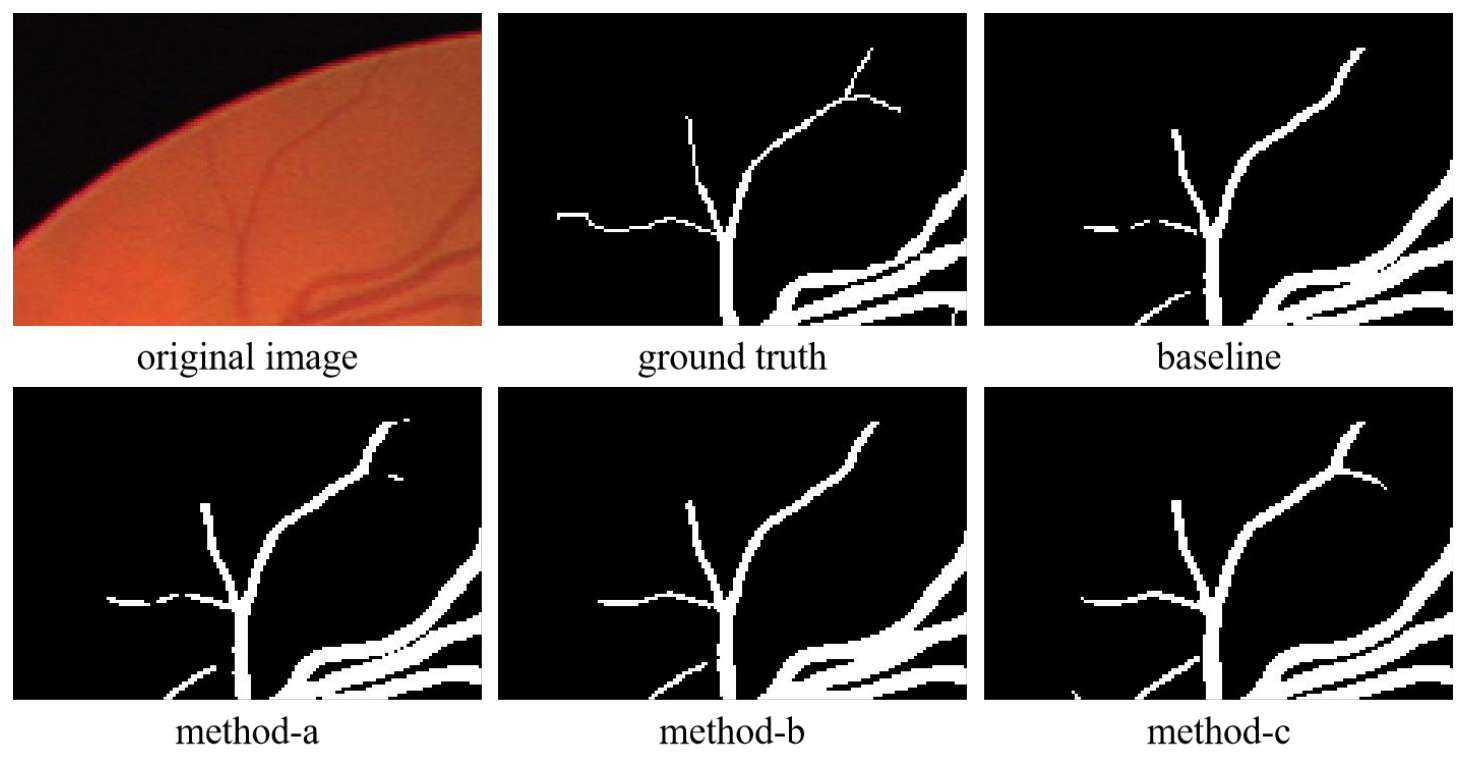}
	\caption{Visualization results of segmented vessels for different network configurations in the ablation study.}
	\label{fig:experiment}
	\vspace{-0.3cm}
\end{figure*}

We perform the ablation study to check whether the proposed semantics-guided module and recursive refinement are 
effective or not. This study is performed by evaluating different network configurations on the DRIVE dataset. 
Detailed results are summarized in Table~\ref{T1}. 
The baseline method is only a 3-layered U-Net trained by deep supervision, 
when neither the semantics-guided module nor the recursive refinement is used. 
The method-a and method-b respectively denote that either the semantics module or the recursive refinement 
is utilized, while the method-c denotes that both of them takes effect. 
Just from the three segmentation related metrics, the four methods are not different too much, though 
the method-c has achieved the highest values on AUC and SE. However, they quite differ in the 
metrics related to vessel connectivity. It can be seen that both INF and COR become better when 
either the semantics-guide module or recursive refinement is activated. The best performance 
for vessel connectivity is achieved when both of them take effect. Compared with the baseline method, 
the INF is decreased by 21.7\% while the COR is increased by 12.2\% for the method-c. Therefore, 
these results quantitatively demonstrate that the connectivity of segmented vessels can be enhanced by 
using the proposed recursive semantics-guided network. 

Besides, we also give the visualization results 
in Fig.~\ref{fig:experiment} to show how these methods are visually different on binary segmentation results. 
Compared with the baseline, the semantics-guided module improves the connectivity slightly 
and makes more capillary vessels be extracted. The refinement strategy can significantly 
avoid the breakpoints in segmented vessels. By using the semantics-guided module together with the refinement, 
not only the vessel extraction becomes more accurate but also the connectivity is largely enhanced.

\subsection{Comparison with Other Leading Methods}

\begin{figure*}[!t]
	\centering
	\includegraphics[width=1.0 \linewidth]{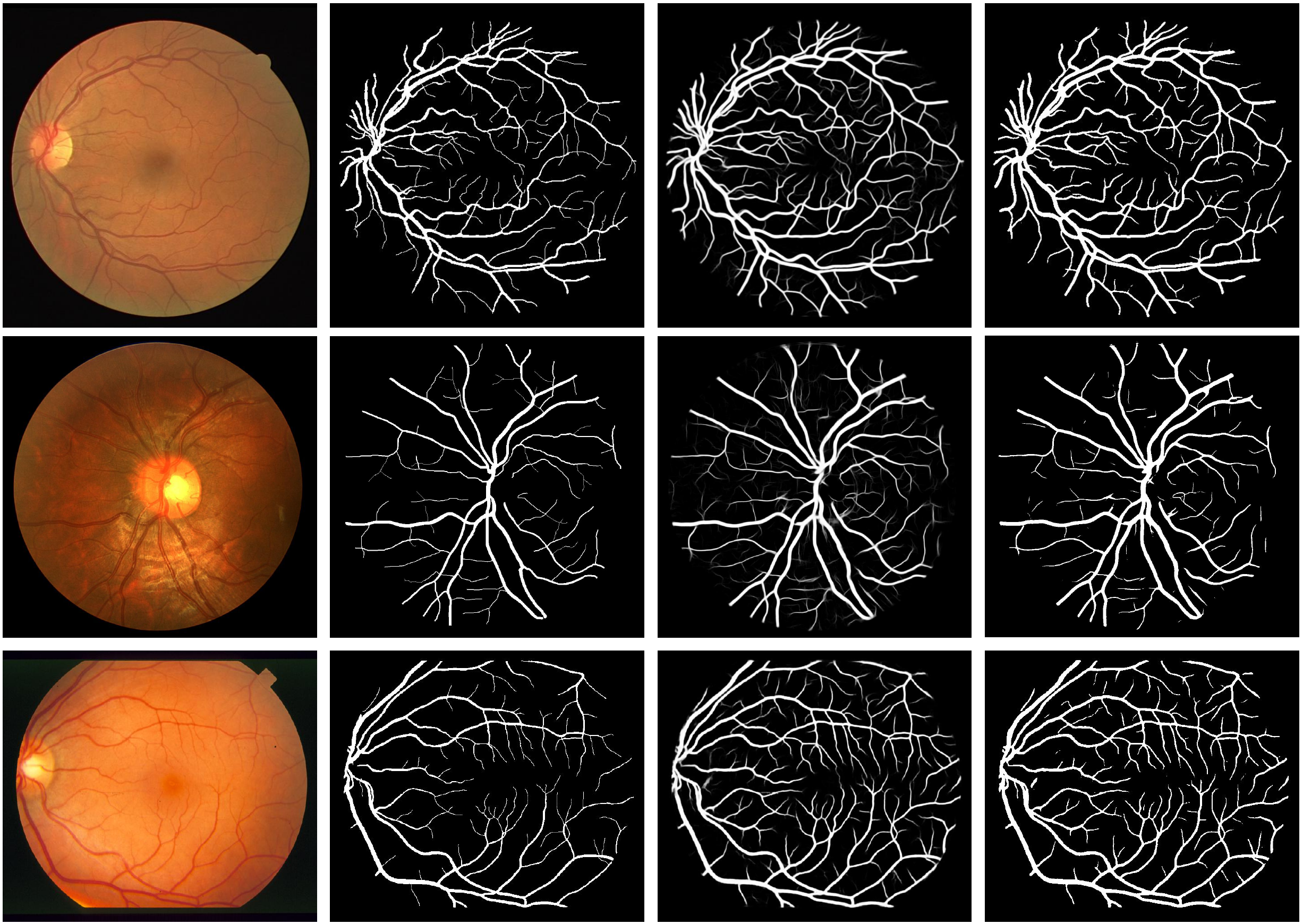}
	\caption{Retinal vessel segmentation results for the proposed method on the DRIVE, CHASE\_DB1 and STARE datasets. From the column in left to right, the retinal images, the ground truth, the predicted probability maps and the corresponding binary segmentation of the propose method are given respectively.}
	\label{fig:result}
	\vspace{-0.2cm}
\end{figure*}

We compare the proposed method with three leading methods by evaluating them on the DRIVE, CHASE\_DB1 
and STARE datasets. Ara \textit{et al.} \cite{araujo2019deep} propose a stacked variational auto-encoder based network to 
improve the connectivity of segmented vessels, which could be the only work that is targeted for boosting 
vessel connectivity and published in recent years, according to our knowledge. The other two methods are specifically aimed 
for improving the retinal vessel segmentation. Oliveira \textit{et al.}~\cite{oliveira2018retinal} propose a
multiscale fully convolutional network for vessels segmentation, which combines the multiscale analysis provided by the 
stationary wavelet transform. Xu \textit{et al.}~\cite{Xu2020} improve vessel segmentation by introducing a carefully 
designed semantics and multi-scale aggregation network. The results are summarized in Table~\ref{T2}. 
If only evaluating these methods by the AUC, SE and SP, one might consider that the performance of 
our method were roughly equal to the others. However, the difference becomes obvious when they are 
compared by using the INF and COR, which can quantify the connectivity of segmented vessels. Especially for 
the CHASE\_DB1 dataset, our method can decrease the INF by 19.1\% while increase the COR by 15.4\%, compared 
with the 2nd ranking method. These results show that the proposed method outperforms the three leading methods and 
achieves the best performance on vessel connectivity. 
Finally, we give examples of our method for retinal vessel segmentation on the three public 
datasets in Fig.~\ref{fig:result}.

\begin{table}[!t]
	\caption{Comparison of the proposed method with other leading methods for retinal vessel segmentation. (A smaller value of INF means better performance.)}
	\label{T2}
	\centering
	\includegraphics[width=1.0\linewidth]{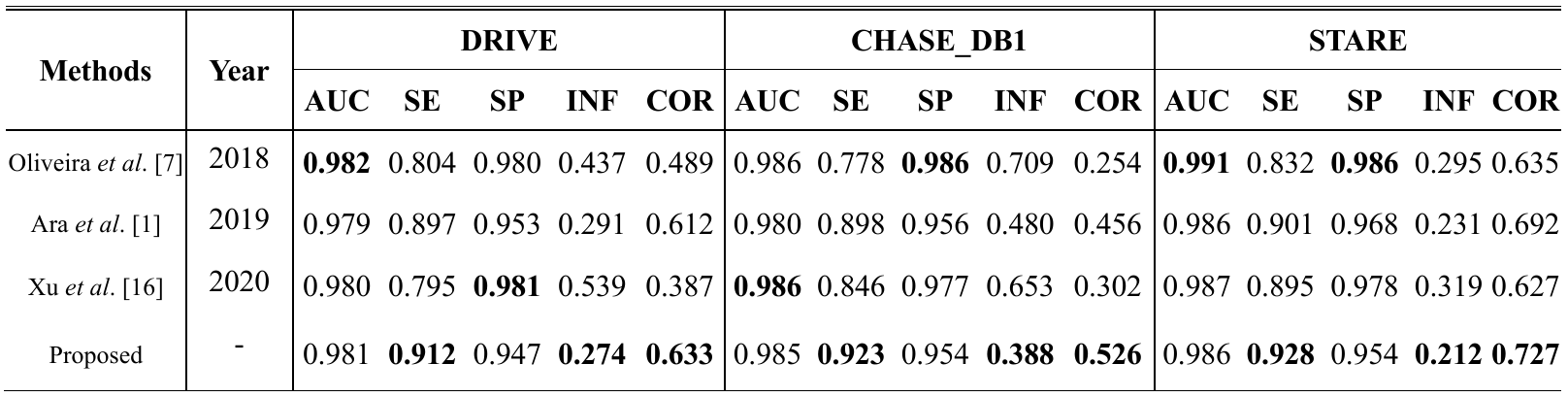}
	\vspace{-0.5cm}
\end{table}

\section{Conclusion}
In this paper, we propose a recursive semantics-guided network for better connectivity on retinal vessel segmentation. 
It is featured by a semantics-guided module that can fully exploit semantics information for guiding the network 
to learn more powerful features, and a recursive refinement that can iteratively enhance the results while saving 
network parameters. Its efficiency is demonstrated from extensive experimental results.

\bibliographystyle{splncs04}
\bibliography{refs}

\end{document}